\documentclass[epjCONF, onecolumn]{svjour}
\usepackage{graphicx}
\usepackage[varg]{txfonts} 
\usepackage[latin1]{inputenc}
\session-title{Assembling the puzzle of the Milky Way}
\begin{document}
\title{Molecular gas and star formation in the Milky Way}
\author{Francoise Combes\thanks{\email{francoise.combes@obspm.fr}} }
\institute{Observatoire de Paris, LERMA et CNRS, 61 Av de l'Observatoire, F-75014 Paris, France}
\abstract{
The dense molecular gas is the ideal tracer of the spiral structure
in the Milky Way, and should be used intensively to solve the puzzle of
its structure. In spite of our position inside the plane, we can hope to disentangle
the structures, with position-velocity diagrams, in addition to $(l-b)$. I summarize
the state of the art simulations of gas flows in the MW, and describe what can be done
to improve the models, taking into account the star formation, in view of what
is already done in external galaxies, with a more favorable viewing angle.
} 
\maketitle

\section{The importance of the gas}
\label{intro}

In the near future, GAIA will make a breakthrough in our knowledge of the 
stellar component of our Galaxy and its dynamics. But the knowledge in its
gas component may not progress as much.
Reproducing the gaseous spiral structure of our Galaxy is a difficult enterprise,
given our internal position. Distances of the various features or objects are derived through a
kinematical model, with near-far ambiguities inside the solar circle. 
The first deprojections of the Galactic plane in the atomic gas
observed at 21cm (Oort et al. 1958) had only a very sketchy and approximative spiral
morphology, without identifying the actual arms, and their continuity.
One of the first successful models
was that from Georgelin \& Georgelin (1976) of four tightly-wound arms, traced by OB associations, optical or
radio HII regions, or molecular clouds. As in other galaxies, the spiral structure
is better contrasted in the gas, atomic 
(HI, Liszt \& Burton 1980) and molecular (CO surveys, Dame et al 2001), 
because of its low velocity dispersion,
and its confinement to the plane.

\begin{figure}
\centering
 \includegraphics[angle=-0,width=7cm]{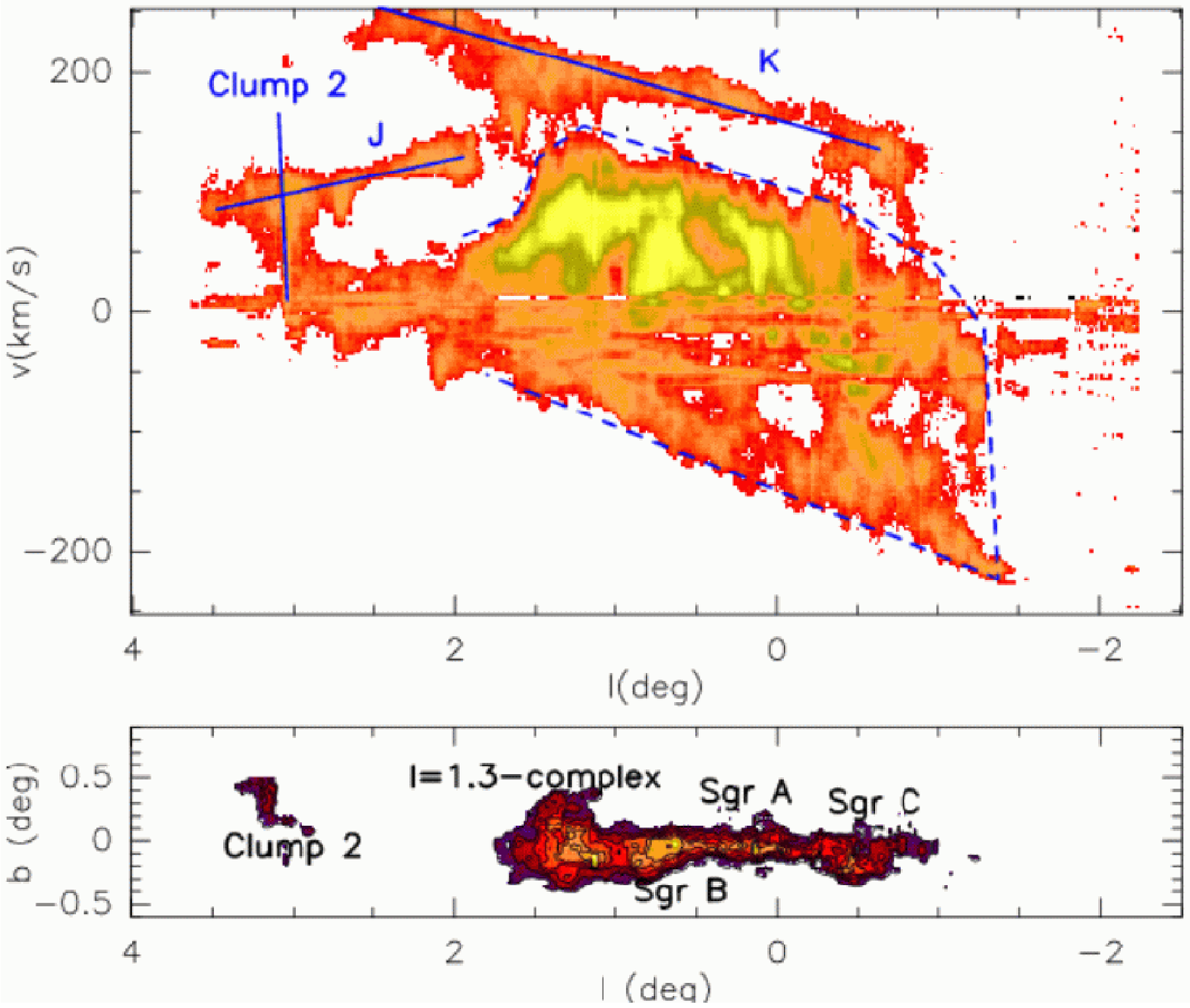} 
 \includegraphics[angle=-0,width=4.1cm]{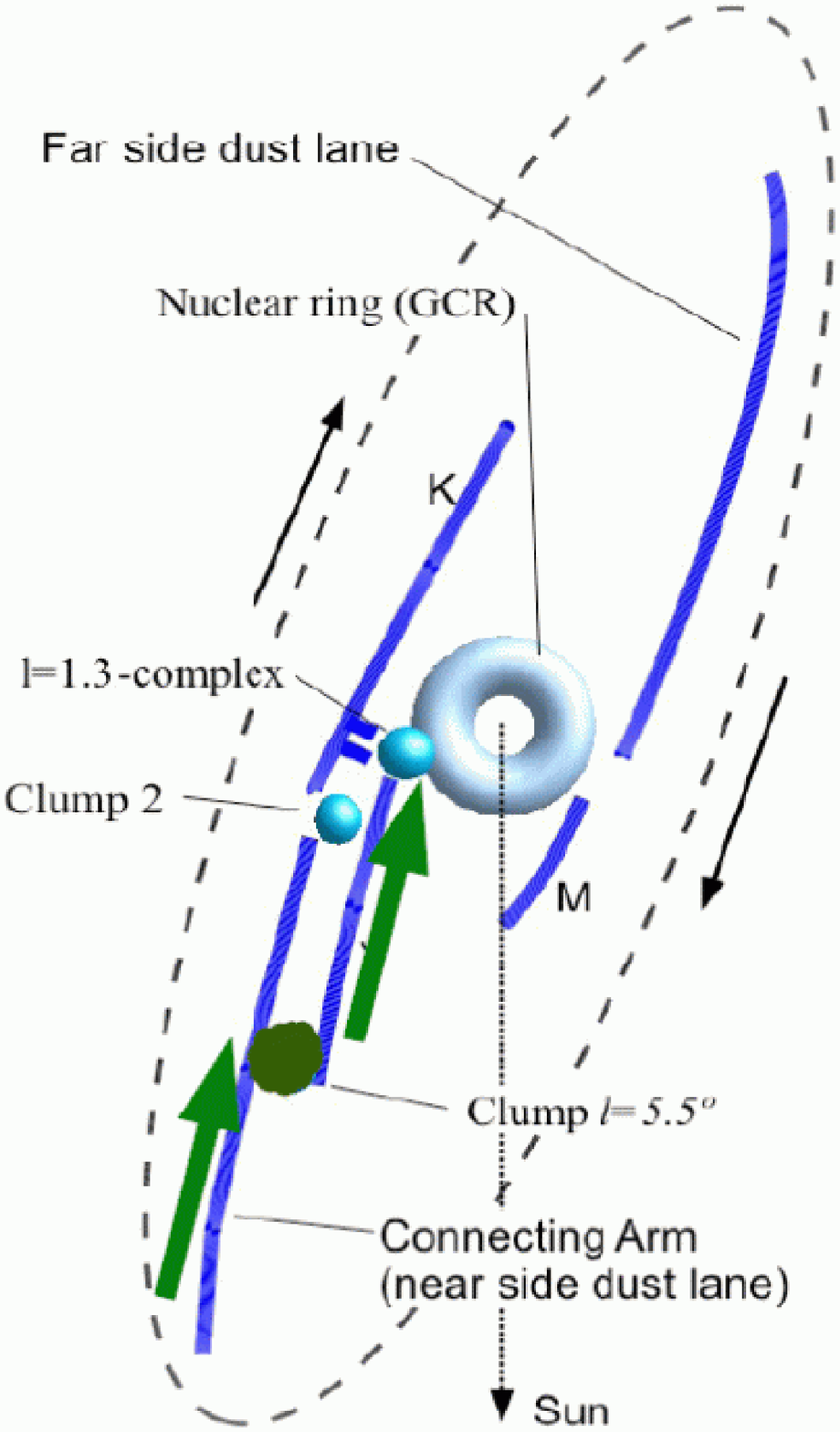} 
\caption{{\bf Left} Longitude-velocity diagram of the CO(1-0) emission in the CMZ (Central Molecular Zone, 
Bally et al 1988).  Some remarkable features are underlined, as Clump2, or strucrures K, J (Rodriguez-Fernandez 
et al 2006). At the bottom is the integrated intensity map.
{\bf Right} Schematic reconstruction in the galactic plane of some of the remarkable structures,
inside the delineated bar (Rodriguez-Fernandez \& Combes 2008).  }
\label{fig1} 
\end{figure}

Position-velocity (P-V) diagrams are particularly instructive, revealing the high velocity ($\sim$560km/s)
Central Molecular Zone (CMZ) near zero longitude, with a molecular ring, connecting arm, 3kpc arm, etc.
(cf Figure \ref{fig1}).
 The existence of a bar has long been suspected from non-circular motions towards the center, and has been
directly confirmed by COBE and 2MASS (e.g. Lopez-Corredoira et al 2005). Near-infrared images show clearly
the peanut bulge, which is thought to be formed through vertical resonance with the bar (e.g. Combes et al
1990).  The CMZ has a peculiar parallelogram shape in P-V diagram (Bally et al 1988), that has been first
interpreted in terms of cusped x1 and almost circular x2 periodic orbits, and associated gas flows, by Binney et al (1991).
 Then Fux (1999) carried out fully self-consistent N-body and hydrodynamical simulations of
 stars and gas to form a barred spiral, and fit the Milky Way. He succeeded remarkably to reproduce
the HI and CO P-V diagrams with a bar pattern speed of about 40 km/s/kpc,
implying a corotation at 5kpc, and an ILR producing the x2 orbit inside 1kpc radius. The spiral structure
has essentially 2 arms starting at the end of the bar.
  
\section{Several possible models}
 More recently, new efforts to reconstitute the spiral structure in the galactic plane have been attempted in the HI gas
(Levine et al 2006), and in the CO gas (Nakanishi \& Sofue 2006, Englmaier et al 2009).
 The best fit could be two arms, starting at the end of the bar, with a pitch angle
of 12$^\circ$, although four arms are still possible. Pohl et al (2008) have tried novel deprojections,
by simulating the gas flow with SPH in a bar potential, and obtaining distances with a kinematical model
derived from the non-circular velocity field obtained.  A test of the procedure with a 2 arms+bar
fiducial model, with only one pattern speed, retrieves after deprojection a four arms spiral. 

 It is frequent in external galaxies (and in simulations) that several pattern speeds develop in 
a disk, and in particular the spiral could rotate slower than the bar. This has been explored
by Bissantz et al (2003), who find as a best fit $\Omega_p$= 60km/s/kpc and 20km/s/kpc for the bar 
and spiral respectively. They simulate gas flows in a fixed potential, and find that fixing the spiral
potential in addition to the bar gives better fits.  

In the 2MASS stellar counts, a  nuclear bar has been found by Alard (2001),
and a CO nuclear bar corresponds (Sawada et al 2004). New simulations
of gas flow in a two-bar models have been done by Rodriguez-Fernandez \& Combes (2008),
who find a best fit when the two bars are nearly perpendicular, and the bar-spiral pattern is
about 35 km/s/kpc (similar to Fux, 1999).  The model shows the far-side symmetric of the 3kpc arm,
which has just been discovered in the CO P-V diagram (Dame \& Thaddeus 2008).
It reproduces also the connecting arm (characteristic leading dust lanes along the bar, fig \ref{fig1}).
No evidence is found of lopsidedness in the stellar potential, and the CO lopsidedness must
be a purely gaseous phenomenon.
Other prominent features have not yet been interpreted, such as the warp or tilt of the 
nuclear gas structure.
Baba et al (2010) have included in their model of the Milky Way more detailed physics,
in particular the multi-phase interstellar medium, its self-gravity, star-formation and supernovae feedback. 
This allows them to reproduce the clumpy morphology observed in the P-V diagrams of CO emission. 

It is interesting to note that the Galaxy is finally more symmetric
than previously thought.
Dame et al (2011) have discovered through CO emission a spiral feature in the distant outer Galaxy
 in the first quadrant, as a continuation of the Scutum-Centaurus arm.

\begin{figure}
\centering
 \includegraphics[angle=-0,width=10cm]{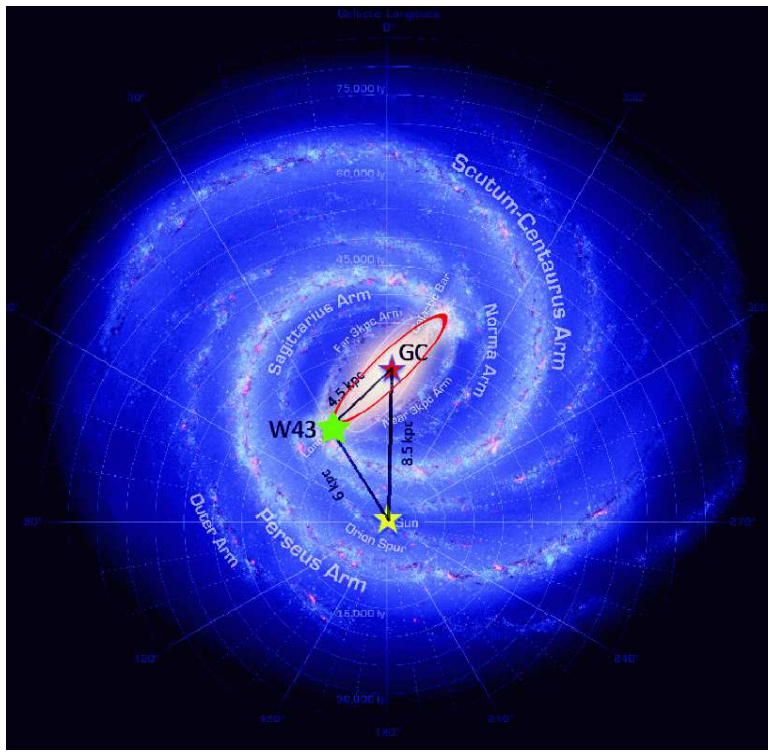} 
\caption{Artist view of the Milky Way with the long bar outlined by a red ellipse
(Churchwell et al. 2009). W43 is located at the extremity of the bar.
From Nguyen-Luong et al (2011).}
\label{fig2} 
\end{figure}

\section{Masers and star formation}

A very performant method to determine precisely the distances has been exploited recently: the 12 GHz methanol masers
 associated to High Mass Star Formation Regions (HMSFR), and their trigonometric parallax determined
from VLBI (see e.g. 
the Bessel survey, Brunthaler et al 2011). A spectacular example has been studied toward
the HMSFR G9.62+0.20 by Sanna et al (2009). They determine a distance of D= 5.2 $\pm$0.6 kpc, while
with a LSR velocity of 2km/s, the kinematical distances are 0.5 and 16kpc!
The region has a large non-circular motion, composed of a radial peculiar velocity 
of 41km/s outward, and an azimuthal counter-rotation of 60km/s.
When considering that the G9.62+0.20 massive star forming region is part
of the near side 3kpc arm, then the kinematic peculiarites are easily
explained with the bar.

Reid et al (2009) combined the trigonometric parallaxes and the proper
motions of the HMSFR masers to determine the morphology of local spiral arms, and their
pitch angle.  They find that star-forming regions on average are orbiting 
the Galaxy with a lag of 15 km/s with respect to circular orbits 
(17km/s for 6.7 GHz masers, Rygl et al 2010). Kinematic distance are sometimes
wrong by a factor 2.
The scatter of HMSFR locations within the arm is larger than parallax errors,
and limit the determination of the arm pitch angles. The data are precise
enough to imply a redetermination of galactic main parameters,
solar radius or velocity (McMillan \& Binney 2010, Schoenrich et al 2010).

\smallskip
From large-scale surveys of gas emission (HI, CO) but also dust at many wavelegnths, it is possible to identify
star formation regions, and their location in the spiral structure morphology. 
Nguyen-Luong et al (2011) have identified W43 as a large ($\sim$140 pc) and coherent 
complex of molecular clouds that is surrounded by an atomic gas envelope ($\sim$290 pc). W43 is particularly
massive (7 10$^6$ M$_\odot$) and concentrated. Its position is conspicuous, at the end of the bar, where
spiral arms begin to wind out (cf Figure \ref{fig2}).
HII Regions are frequently located there in barred galaxies, at the crossing of several orbital streams.
Its star formation rate is up to 0.1 M$_\odot$/yr, and its star formation efficiency SFE= 25\%/Myr.

\section{Conclusions}

For the gaseous medium, it is  presently
still difficult to disentangle distances and dynamical effects.
While stellar dynamics will make considerable progress
with GAIA, we need the distances and proper motions in the interstellar medium.
The high mass star formation masers method, determination of
parallaxes and propoer motions with VLBI, is highly promising.

The spiral structure of our Galaxy has the largest contrast in the gas.
The number of arms and their precise locations are not yet solved,
the spiral structure might include branchings and harmonics, and be
more complex than the stellar one, with two arms and a bar.
The central structure (central molecular zone) is
 lopsided, and may be tilted and warped. The structure should
be studied in more details, and could be the consequence of gas accretion.
The number of patterns, and their respective pattern speeds, is still
not well known territory. From near-infrared surveys,
it is likely that a secondary bar is embedded in the primary bar.

\end{document}